%% file: template.tex
\documentclass{article}

\usepackage{arxiv}

\usepackage[utf8]{inputenc} 
\usepackage[T1]{fontenc}    
\usepackage{hyperref}       
\usepackage{url}            
\usepackage{booktabs}       
\usepackage{amsfonts}       
\usepackage{nicefrac}       
\usepackage{microtype}      
\usepackage{lipsum}		
\usepackage{graphicx}
\usepackage{natbib}
\usepackage{doi}
\usepackage{enumerate}
\usepackage{amsmath}
\usepackage{amssymb}
\usepackage{cleveref}
\usepackage{xcolor}
\usepackage{acronym}
\usepackage{multirow}
\newcommand{\y}{\mathbf{y}}
\newcommand{\Y}{\mathbf{Y}}
\newcommand{\x}{\mathbf{x}}
\newcommand{\X}{\mathbf{X}}

\newcommand{\Z}{\mathbf{Z}}

\newcommand{\Kk}{\mathbf{K}}
\newcommand{\Lk}{\mathbf{L}}

\newcommand\indep{\protect\mathpalette{\protect\independenT}{\perp}}
\def\independenT#1#2{\mathrel{\rlap{$#1#2$}\mkern2mu{#1#2}}}
\newcommand{\R}{\mathbb{R}}
\newcommand{\mmd}{\textsc{mmd}}
\renewcommand{\P}{\mathbb{P}}
\renewcommand{\H}{\mathcal{H}}
\newcommand{\G}{\mathcal{G}}
\newcommand{\hsic}{\textsc{hsic}}

\newacro{mmd}[\textsc{mmd}]{maximum mean discrepancy}
\newacro{hsic}[\textsc{hsic}]{Hilbert-Schmidt independence criterion}
\newacro{rkhs}[\textsc{rkhs}]{reproducing kernel Hilbert space}
\newacro{rff}[\textsc{rff}]{random Fourier feature}
\newacro{sdg}[\textsc{sdg}]{Sustainable Development Goal}

\title{Kernel Two-Sample and Independence Tests for Non-Stationary Random Processes}

\author{Felix Laumann \\
 Department of Mathematics, Imperial College London, SW7 2AZ, United Kingdom \\
 \texttt{f.laumann18@imperial.ac.uk}
 \And Julius von K\"ugelgen \\
 MPI for Intelligent Systems, Max-Planck-Ring 4, 72076 T\"ubingen, Germany \\
 Department of Engineering, University of Cambridge, CB2 1TN, United Kingdom 
 \And Mauricio Barahona \\
 Department of Mathematics, Imperial College London, SW7 2AZ, United Kingdom}

\begin{document}
\maketitle

\begin{abstract}
    Two-sample and independence tests with the kernel-based \ac{mmd} and \ac{hsic} have shown remarkable results on i.i.d.\ data and stationary random processes. However, these statistics are not directly applicable to non-stationary random processes, a prevalent form of data in many scientific disciplines. In this work, we extend the application of \ac{mmd} and \ac{hsic} to non-stationary settings by assuming access to independent realisations of the underlying random process.
    These realisations---in the form of non-stationary time-series measured on the same temporal grid---can then be viewed as i.i.d.\ samples from a multivariate probability distribution, to which \ac{mmd} and \ac{hsic} can be applied.
    We further show how to choose suitable kernels over these high-dimensional spaces by maximising the estimated test power with respect to the kernel hyper-parameters.
    In experiments on synthetic data, we demonstrate superior performance of our proposed approaches in terms of test power when compared to current state-of-the-art functional or multivariate two-sample and independence tests.
    Finally, we employ our methods on a real socio-economic dataset as an example application. 
\end{abstract}

\input{Section_1}

\input{Section_2}

\input{Section_3}

\input{Section_4}

\input{Section_5}

\newpage
\bibliographystyle{unsrtnat}
\bibliography{references} 

\newpage 
\appendix

\section{Appendix}


\subsection{Results for realisations with varying number of time points, \textit{T}} \label{results_coarse}

\subsubsection{\ac{mmd}} We show here the results for mean and variance shifts for $m = n = 100$, but the results are similar for all tested sample sizes $m = n = 100, 200, 300, 500$,

\begin{figure}[h]
    \centering
    \begin{minipage}{.47\textwidth}
    \includegraphics[width=\textwidth]{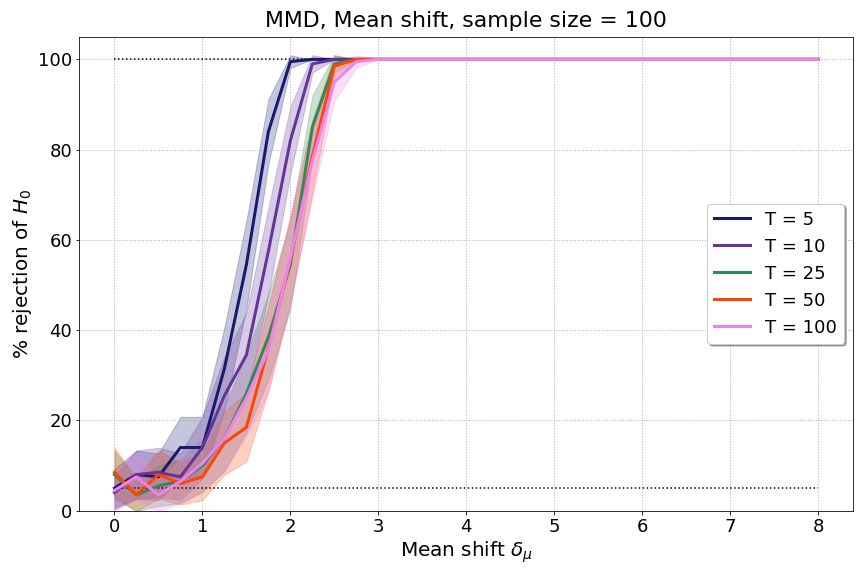}
    \end{minipage}
    \hspace{0.5cm}
    \begin{minipage}{.47\textwidth}
    \includegraphics[width=\textwidth]{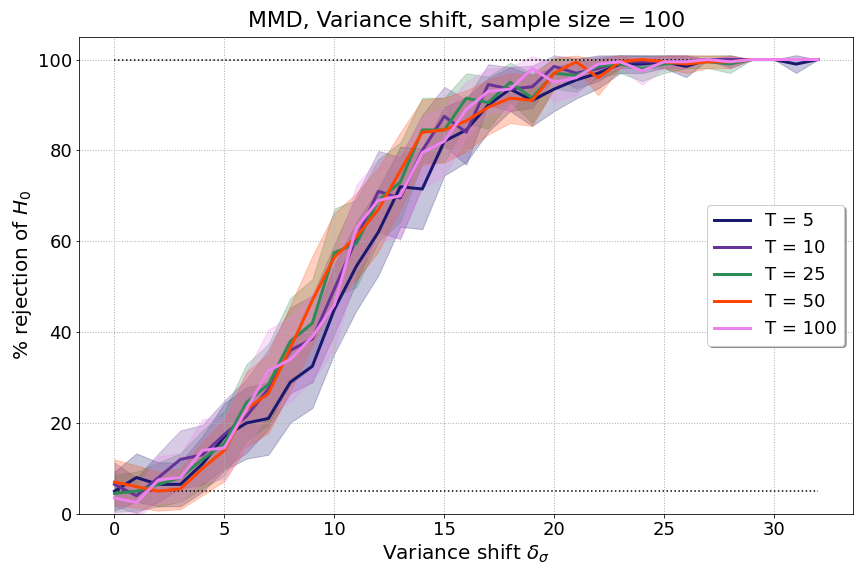}
    \end{minipage}
    \caption{Results of \ac{mmd}-based homogeneity test with $T = [5, 10, 25, 50, 100]$: Percentage of rejected $H_0$ for mean shift \textit{(left)} and variance shift \textit{(right)} for sample sizes $m = n = 100$ and $T$ discrete time points in $d=1$ dimensions.}
    \label{fig:mean_var_shift_all_T}
\end{figure}

\subsubsection{\ac{hsic}} Experiments for linear dependence and dependence through shared second basis function coefficient for various $T$. We find that the granularity of measurements over time does not influence the text power significantly.

\begin{figure}[h!]
    \centering
    \begin{minipage}{.47\textwidth}
    \includegraphics[width=\textwidth]{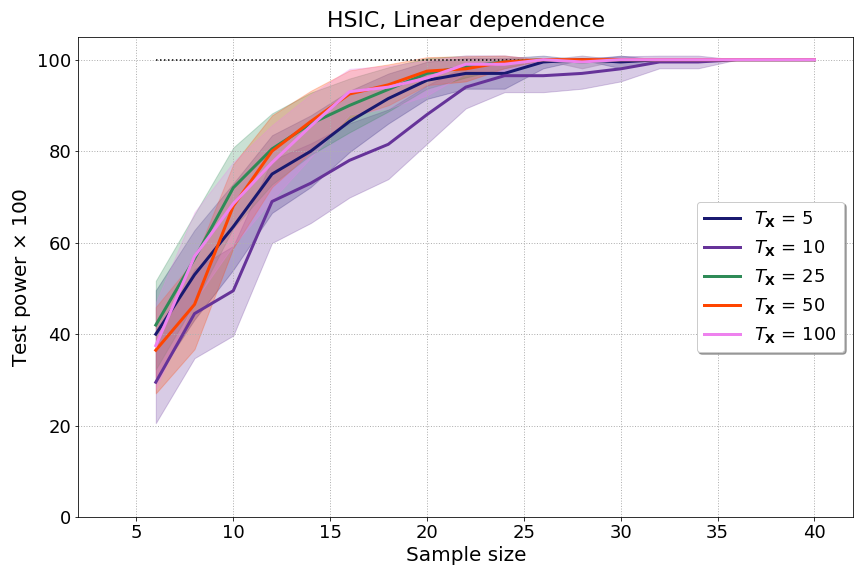}
    \end{minipage}
    \hspace{0.5cm}
    \begin{minipage}{.47\textwidth}
    \includegraphics[width=\textwidth]{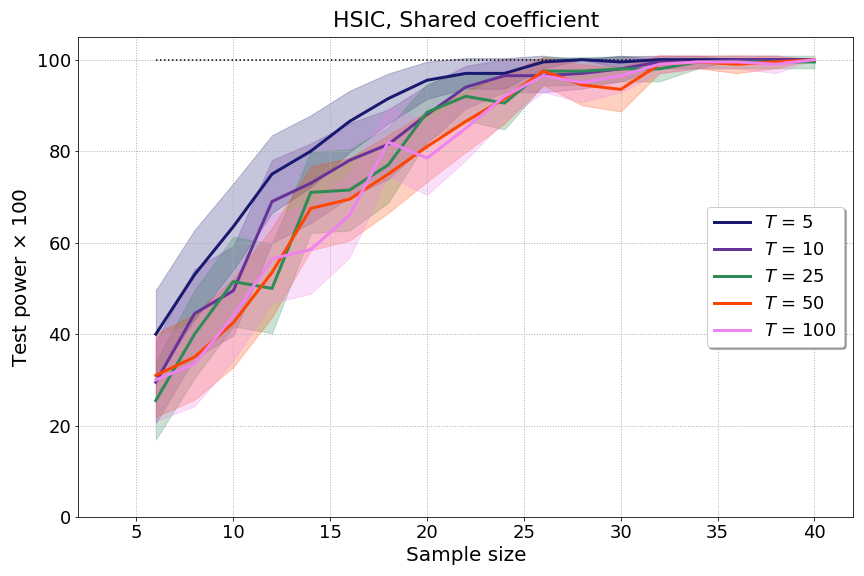}
    \end{minipage}
    \caption{Results of the \ac{hsic}-based independence test: Test power for linear dependence and dependence through shared coefficient as sample size is varied for various numbers of time points $T = [5, 10, 25, 50, 100]$.}
    \label{fig:dep_linear_higher_T}
\end{figure}

\newpage 
\subsection{Test power maximisation} \label{delta_search}
\subsubsection{\ac{mmd}} For mean shift experiments for \acs{mmd}, we pre-define a linear search space with $11$ values for the Gaussian kernel bandwidth $\sigma$ due to the dependence on $\delta_{\mu}$, and similarly for variance shift experiments (both stated in Table~\ref{search_spaces_MMD}). These search spaces resulted from extensive manual explorations for all shifts and sample sizes. We acknowledge that the test power may be further improved with search spaces of finer granularity.

\begin{table}[h!]
\centering
\caption{Linear search spaces for bandwidth $\sigma$ in \acs{mmd} mean \textit{(left)} and variance \textit{(right)} shift experiments.}
\label{search_spaces_MMD}
\scalebox{0.85}{
\begin{minipage}{0.45\linewidth}
\centering
\begin{tabular}{l cccc}
\toprule
$\delta_\mu$ & 0 – 2    & 2.25 – 3    & 3.25 – 5    & 5.5 – 8         \\[0.1cm]
\empty & \multicolumn{3}{c}{step size = 0.25} & step size = 0.5 \\
\midrule
\empty & 1        & 6           & 11          & 16              \\
\empty & 3        & 8           & 13          & 18              \\
{\multirow{5}{*}{\rotatebox[origin=c]{90}{search space for $\sigma$}}} & 5        & 10          & 15          & 20              \\
\empty & 7        & 12          & 17          & 22              \\
\empty & 9        & 14          & 19          & 24              \\
\empty & 11       & 16          & 21          & 26              \\
\empty & 13       & 18          & 23          & 28              \\
\empty & 15       & 20          & 25          & 30              \\
\empty & 17       & 22          & 27          & 32              \\
\empty & 19       & 24          & 29          & 34              \\
\empty & 21       & 26          & 31          & 36             \\
\bottomrule
\end{tabular}
\end{minipage}
\hspace{2cm}
\begin{minipage}{0.45\linewidth}
\centering
\begin{tabular}{l ccc}
\toprule
$\delta_\sigma$ & 0 – 4    & 5 – 14    & 15 – 32    \\[0.1cm]
\empty & \multicolumn{3}{c}{step size = 1} \\
 \midrule
\empty & 10       & 20        & 30         \\
\empty & 12       & 22        & 32         \\
{\multirow{5}{*}{\rotatebox[origin=c]{90}{search space for $\sigma$}}} & 14       & 24        & 34         \\
\empty & 16       & 26        & 36         \\
\empty & 18       & 28        & 38         \\
\empty & 20       & 30        & 40         \\
\empty & 22       & 32        & 42         \\
\empty & 24       & 34        & 44         \\
\empty & 26       & 36        & 46         \\
\empty & 28       & 38        & 48         \\
\empty & 30       & 40        & 50        \\
 \bottomrule
\end{tabular}
\end{minipage}
}
\end{table}


\subsubsection{\ac{hsic}} We define search intervals of both $\sigma_\X$ and $\sigma_\Y$ across all angles $\theta$, but different for the student-t, uniform, and exponential distributions. For student-t and exponential distributions, both $\sigma_\X$ and $\sigma_\Y$ were chosen as $20$ evenly spaced numbers on a linear scale between $1$ and $20$. For uniform distributions, both $\sigma_\X$ and $\sigma_\Y$ were chosen as $40$ evenly spaced numbers on a linear scale between $1$ and $40$. These search spaces resulted from extensive manual explorations for all angles and distributions.
We acknowledge that the test power may be further improved with search spaces of finer granularity.

\newpage
\subsection{Distribution specifications for basis function coefficients in rotation mixing} \label{tab:dist_coeff}

\begin{table}[h!]
\centering
\caption{Specifications of distributions for the rotation mixing. They are a subset of the distributions in \cite[Table 3]{gretton2005kernel}, and $\Z$ is a proxy for both $\X$ and $\Y$.}
    \vspace{0pt}
    \centering
    \scalebox{0.85}{
    \begin{tabular}{l c c}
     \toprule
     Distribution  & 
     \multicolumn{2}{c}{Fourier basis function coefficients} \\ 
     \empty & $\xi_{\Z i1}$ & $\xi_{\Z i2}$ \\
     \midrule
     Exponential & $\lambda = 1.5$ & $\lambda = 3$ \\  
     Student-t & $\nu = 3$ & $\nu = 5$ \\
     Uniform & $\mathcal{U}[-10, 10]$ & $\mathcal{U}[-5, 5]$ \\ 
     \bottomrule
     \end{tabular}}
    \label{distributions_rotation}
\end{table}
\vspace{-0.35cm}
\begin{figure}[h!]
    \centering
    \begin{minipage}{.325\linewidth}
    \centering
        \includegraphics[width=\textwidth]{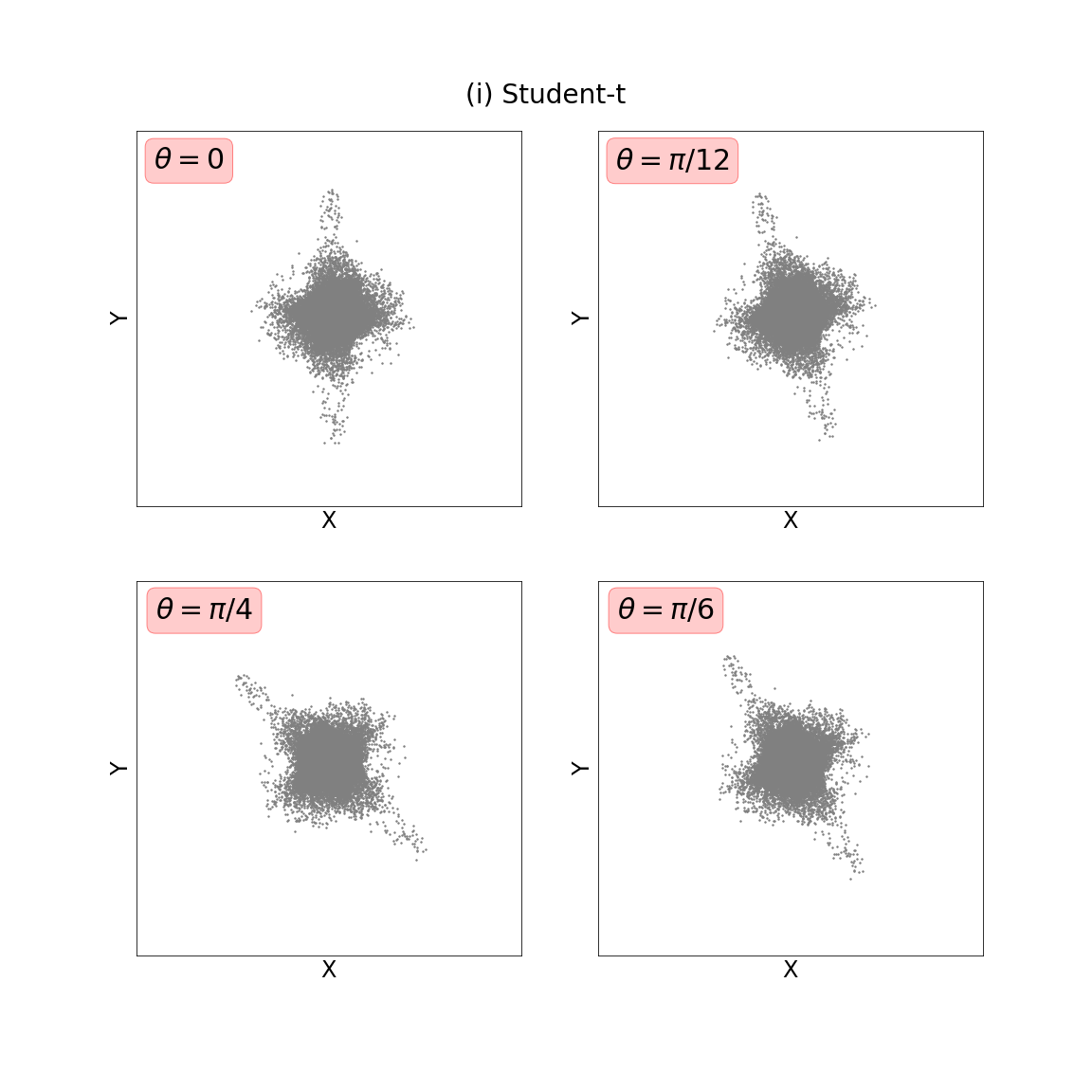}
    \end{minipage}
    \begin{minipage}{.325\linewidth}
    \centering
        \includegraphics[width=\textwidth]{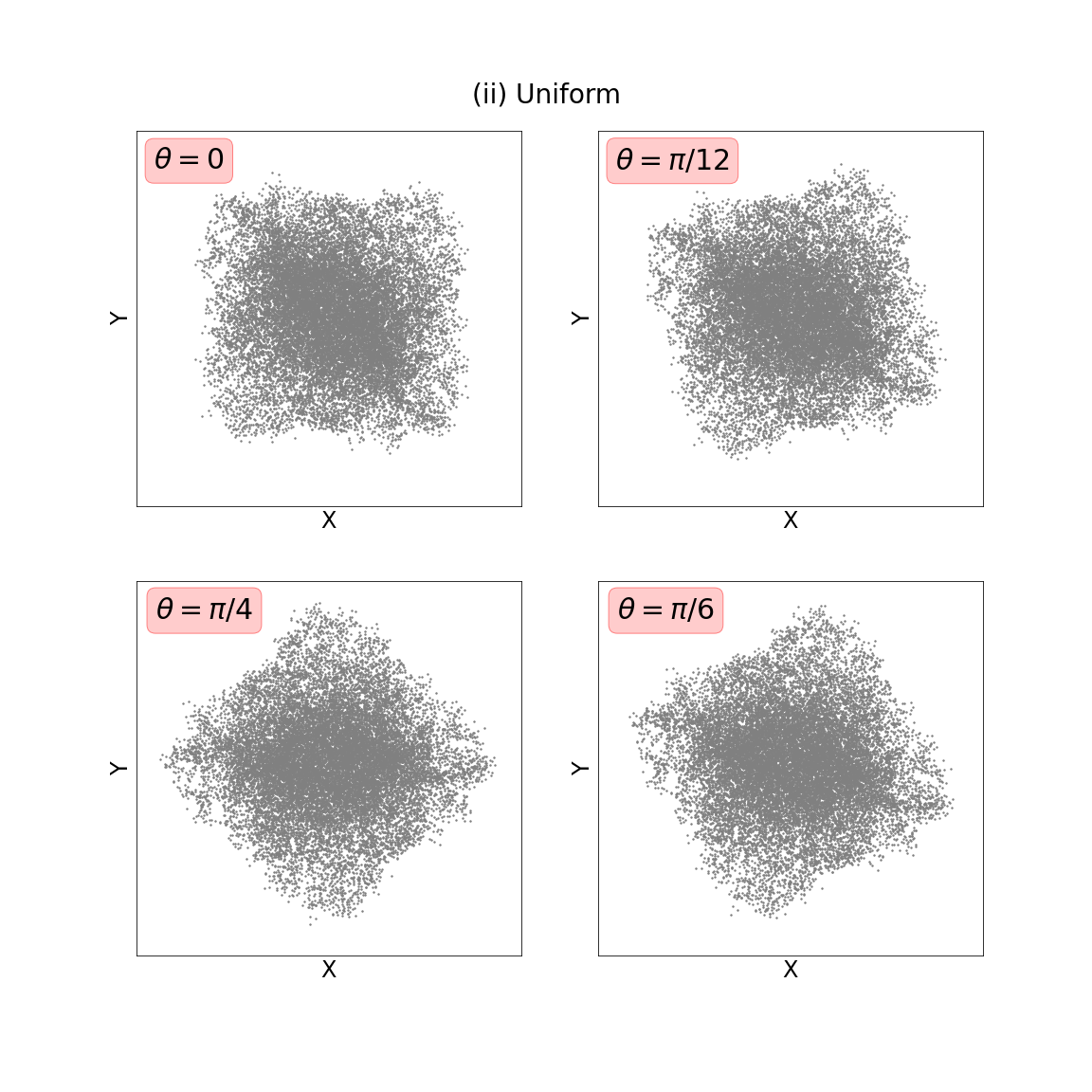}
    \end{minipage}
    \begin{minipage}{.325\linewidth}
    \centering
        \includegraphics[width=\textwidth]{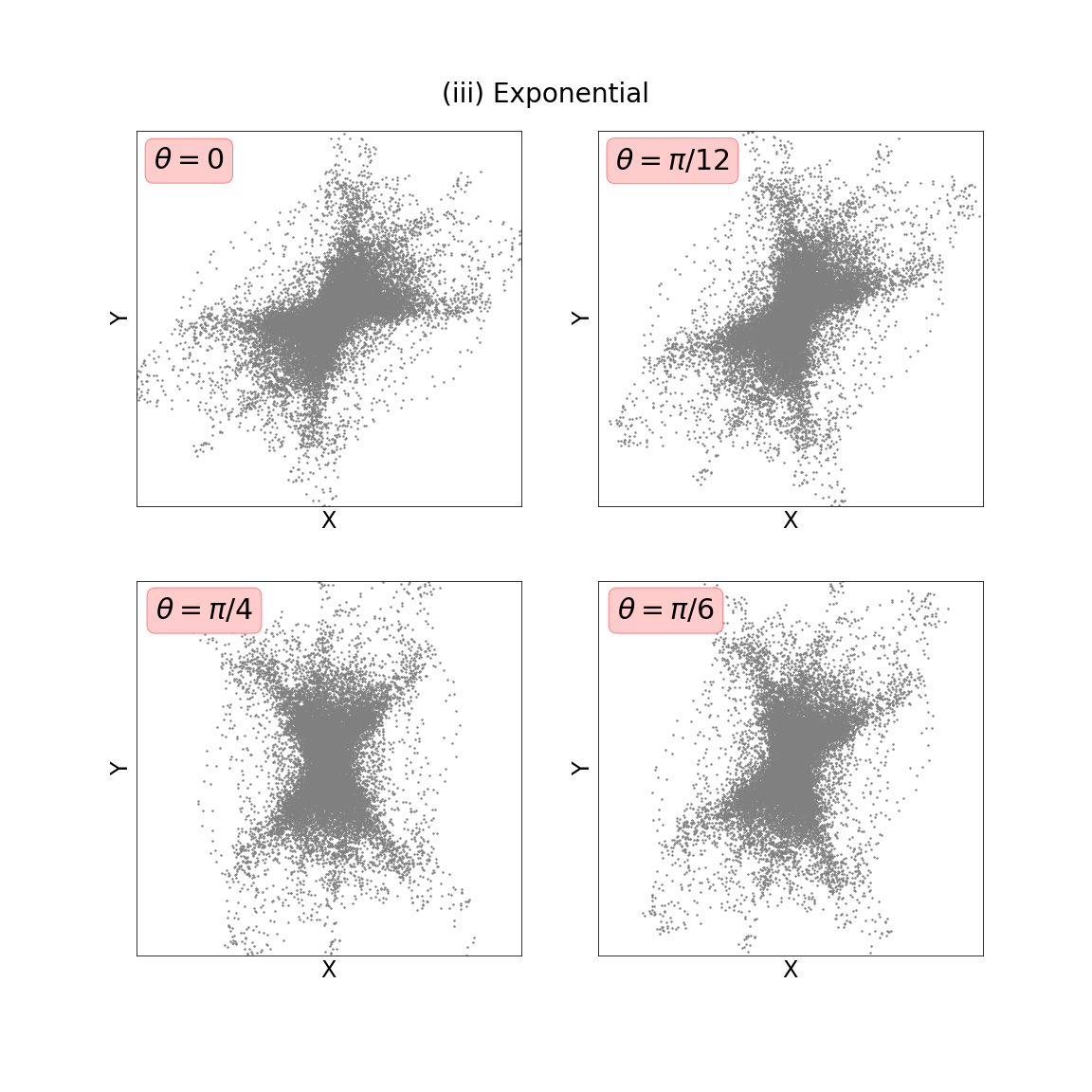}
    \end{minipage}
    \caption{Illustration of $\X$ and $\Y$ with (i) student-t, (ii) uniform, and (iii) exponential basis function coefficients being mixed by different rotation angles $\theta$, ordered clockwise by increasing $\theta$.}
    \label{fig:rotations_illustrations}
\end{figure}

\subsection{SDG dataset} \label{sdg data}
Data of the Indicators measuring the progress of the Targets of the \acp{sdg} can be found at \cite{WBdata}. Each of these Indicators measures the progress towards a specific Target. For instance, an Indicator for Target 1.1, \textit{`by 2030, eradicate extreme poverty for all people everywhere, currently measured as people living on less than \$1.90 a day'}, is the \textit{`proportion of population below the international poverty line, by gender, age, employment status and geographical location (urban/rural)'}. Each of the Targets belongs to one specific Goal (e.g., Target 1.1 belongs to Goal 1, \textit{`end poverty in all its forms everywhere'}). There are $17$ such Goals, which are commonly referred to as the Sustainable Development Goals (\acp{sdg}). We compute averages over all Indicators belonging to one Target for our analyses in Section \ref{sdgs}.

The dataset of \cite{WBdata} has many missing values, especially for the time span 2000-2005. We impute these values using a weighted average across countries (where data is available) with weights inversely proportional to the Euclidean distance between indicators.

\end{document}

%% file: Section_1.tex
\section{Introduction} \label{intro}

Non-stationary processes are the rule rather than the exception in many scientific disciplines such as epidemiology, biology, sociology, economics, or finance. In recent years, there has been a surge of interest in the analysis of problems described by large sets of interrelated variables with few observations over time, often involving complex non-linear and non-stationary behaviours. 
Examples of such problems include the longitudinal spread of obesity in social networks~\cite{christakis2007spread}, disease modelling from time-varying inter- and intra-cellular relationships \cite{barabasi2011network}, behavioural responses to losses of loved ones within social groups~\cite{bond2017complex}, and the linkage between climate change and the global financial system \cite{battiston2017climate}.
All such analyses rely on the statistical assessment of the similarity between, or the relationship amongst, noisy time series that exhibit temporal memory.
Therefore, the ability to test the statistical significance of homogeneity and dependence between random processes that cannot be assumed to be independent and identically distributed (i.i.d.) is of fundamental importance in many fields.

Kernel-based methods provide a popular framework for homogeneity and independence tests by embedding probability distributions in \acp{rkhs} \cite[Section 2.2]{muandet2017kernel}.
Of particular interest are the kernel-based two-sample statistic \acl{mmd} (\ac{mmd}) \cite{gretton2007kernel}, which is used to assess whether two samples were drawn from the same distribution, hence testing for \textit{homogeneity}; and the related \acl{hsic} (\ac{hsic}) \cite{gretton2008kernel}, which is used to assess dependence between two random variables, thus testing for \textit{independence}.
These methods are non-parametric, i.e., they do not make any assumptions on the underlying distribution or the type of dependence.
However, in their original form, both \ac{mmd} and \ac{hsic} assume access to a sample of i.i.d.\ observations---an assumption that is often violated for temporally-dependent data such as random processes.

Extensions of \ac{mmd} and \ac{hsic} to random processes have been proposed \cite{besserve2013statistical,chwialkowski2014wild}. Yet, these methods require the random process to be \textit{stationary}, meaning that its distribution does not change over time.
Whilst it is sometimes possible to approximately achieve stationarity with pre-processing techniques such as (seasonal) differencing or square root and power transformations, such approaches become cumbersome and notoriously difficult, particularly with large sets of variables.
The stationarity assumption can therefore pose severe limitations in many application areas where multiple non-stationary processes must be taken into consideration. When studying the relationships of climate change to the global financial system, for example, factors such as greenhouse gas emissions, stock market indices, government spending, and corporate profits would have to be transformed or assumed to be stationary over time.

In this paper, we show how the kernel-based statistics \ac{mmd} and \ac{hsic} can be applied to \textit{non-stationary} random processes. 
At the heart of our proposed approach is the simple, yet effective idea that realisations of a random process in the form of temporally-dependent measurements (i.e., the observed time series) can be viewed as independent samples from a multivariate probability distribution, provided that they are observed at the same points in time, i.e., over the same temporal grid.
Then, \ac{mmd} and \ac{hsic} can be applied on these distributions to test for homogeneity and independence, respectively.

The remainder of this paper is structured as follows.
After discussing related work in \cref{related_word}, we introduce our applications of two-sample and independence testing with \ac{mmd} and \ac{hsic} to non-stationary random processes in~\cref{method}. We then carry out experiments on multiple synthetic datasets in~\cref{experiments} and demonstrate that the proposed tests have higher power compared with current functional or multivariate two-sample and independence tests under the same conditions. 
We provide an example application of our proposed methods to a socio-economic dataset in~\cref{sdgs} and conclude the paper with a brief discussion in~\cref{conclusion}.

%% file: Section_2.tex
\section{Related work} \label{related_word}

Two-sample and independence tests on stochastic processes have been widely studied in recent years. Under the stationarity assumption, \cite{besserve2013statistical} investigate how the kernel cross-spectral density operator may be used to test for independence, and
\cite{chwialkowski2014wild} formulate a wild bootstrap-based approach for both two-sample and independence tests, which outperforms~\cite{besserve2013statistical} 
in various experiments.
The wild bootstrap in \cite{chwialkowski2014wild} approximates the null hypothesis $H_0$ by assuming there exists a time lag $\tau$ such that a pair of measurements at any point in time $t$, $(x_i, y_i)_t$, is independent of $(x_i, y_i)_{t \pm s}$ for $s\geq\tau$. 
This method is applicable to test for instantaneous homogeneity and independence in stationary processes, but requires further assumptions to investigate non-instantaneous cases: a maximum lag $M \leq \tau$ must be defined as the largest absolute lag for the test. 
This results in multiple hypothesis testing requiring adjustment
by a Bonferroni correction. 
Further, \cite{davis2018applications} have applied \textit{distance correlation} \cite{szekely2007measuring}, a \ac{hsic}-related statistic, to independence testing on stationary random processes.

Beyond the stationarity assumption, two-sample testing in the functional data analysis literature has mostly focused on differences of mean \cite{Horvth2012} or covariance structures \cite{FREMDT2012,Panaretos2010}. However, \cite{pomann2016two} have developed a two-sample test for \textit{distributions} based on generalisations of a finite-dimensional test by utilising functional principal component analysis, and \cite{wynne2020kernel} have derived kernels over functions to be used with \ac{mmd} for the two-sample test. Independence testing for functional data using kernels was recently proposed in \cite{Grecki2018}, but assumes the samples lie on a finite-dimensional subspace of the function space---an assumption not required in our work. Moreover, \cite{zhang2018large} have developed computationally efficient methods to test for independence on high-dimensional distributions and large sample sizes by using eigenvalues of centred kernel matrices to approximate the distribution under the null hypothesis $H_0$ instead of 
simulating a large number of permutations.

%% file: Section_3.tex
\section{\ac{mmd} and \ac{hsic} for non-stationary random processes} \label{method}


\subsection{Notation and assumptions} \label{notation}

Let $\{\X_t\}$ and $\{\Y_t\}$ denote two non-stationary stochastic processes with probability laws $\P_{\X}$ and $\P_{\Y}$, respectively. 
We assume that we observe $m$ independent realisations of $\{\X_t\}$ and $n$ independent realisations of $\{\Y_t\}$ in the form of time series measured at $T_\X$ and $T_\Y$ time points, respectively. 
Said differently, the data samples  $\X = \{\x_i\}_{i=1}^m\overset{\text{i.i.d.}}{\sim}\P_{\X}$ are a set of non-stationary time series, $\x_i = \{x_{i,1}, \dots, x_{i,T_\X}\}$, arriving over the same temporal grid, and similarly for $\Y = \{\y_i\}_{i=1}^n\overset{\text{i.i.d.}}{\sim}\P_{\Y}$ with $\y_i = \{y_{i,1}, \dots, y_{i,T_\Y}\}$.
Note that the measurements $x_{i,t}$ and $y_{i,t}$ are not independent across time.\footnote{We use the terms `sample' and `realisation' interchangeably to denote $\x_i$ and $\y_i$, and use the term 'measurement' to denote the temporally dependent vectors $x_{i,t}$ and $y_{i,t}$.}  

We may view the realisations $\x_i$ and $\y_i$ as samples of multivariate probability distributions of dimension $T_\X$ and $T_\Y$, respectively, which are independent at any given point in time, i.e., $x_{i,t} \indep x_{j,t}$ and $y_{i,t} \indep y_{j,t}$ $\forall t$ and $\forall i\neq j$.
Consequently, we can represent these distributions by their mean embeddings $\mu_\X$ and $\mu_\Y$ in \acp{rkhs}, and use these to conduct kernel-based two-sample and independence tests.
Given a characteristic kernel $k$, i.e., the mean embedding $\mu$ captures all information of a distribution $\P$~\cite{sriperumbudur2010hilbert}, the dependence between measurements in time is captured by the ordering of the variables, and the fact that any characteristic kernel $k$ is injective, thus guaranteeing a unique mapping of any probability distribution into a \acs{rkhs}~\cite{sriperumbudur2011universality}. 

For homogeneity testing ($\P_{\X}\overset{?}{=}\P_\Y$), we use the kernel-based \acs{mmd} statistic and require equal number of measurements $T=T_\X=T_\Y$, 
but allow different sample sizes, $m\neq n$.
For independence testing ($\P_{\X\Y}\overset{?}{=}\P_\X\P_\Y$), we employ the related \acs{hsic}, and in this case number of measurements can differ, but we require the same number of realisations, $m=n$. 
We now describe how two-sample and independence tests can be performed under these assumptions.

\subsection{\ac{mmd} for non-stationary random processes} \label{mmd}

Let $k:\R^{T} \times \R^{T}\rightarrow \R$ be a characteristic kernel, such as the Gaussian kernel $k(x, y) = \exp{(-\|x-y\|^2/\sigma^2)}$, which uniquely maps $\mathbb{P}_\X$ and $\mathbb{P}_\Y$ to their associated \acs{rkhs} $\mathcal{H}_k$ via the mean embeddings $\mu_\X := \int k(\x, \cdot) \, d\mathbb{P}_\X(\x)$ and $\mu_\Y := \int k(\y, \cdot) \, d\mathbb{P}_\Y(\y)$  \cite[Section 2.1]{muandet2017kernel}.
The \ac{mmd} between $\P_\X$ and $\P_\Y$ in $\H_k$ is defined as~\cite{gretton2007kernel}:
\begin{equation}
    \label{eq:population_MMD}
    \mmd^2(\mathcal{H}_k, \mathbb{P}_\X, \mathbb{P}_\Y) := \| \mu_{\X} - \mu_{\Y} \|_{\mathcal{H}_k}^2 \geq 0, \quad \text{with equality iff} \quad \mathbb{P}_{\X} = \mathbb{P}_{\Y}.
\end{equation}
Given samples $\X$ and $\Y$, $\mmd^2 (\mathcal{H}_k, \mathbb{P}_\X, \mathbb{P}_\Y)$ can then be approximated by the following unbiased estimator~\cite{gretton2007kernel}:
\begin{equation}
\label{mmd_u}
        \widehat{\mmd}_u^2(\mathcal{H}_k, \X, \Y) =  \sum_{i=1}^m \sum_{j \neq i}^m \frac{k(\x_i, \x_j)}{m(m-1)} +   \sum_{i=1}^n \sum_{j \neq i}^n \frac{k(\y_i, \y_j)}{n(n-1)} -   2\sum_{i=1}^m \sum_{j=1}^n \frac{k(\x_i, \y_j)}{mn}.
\end{equation}
Henceforth, we drop the implied $\mathcal{H}_k$ for ease of notation. 

Using $\widehat{\mmd}_u^2(\mathbf{X}, \mathbf{Y})$ as a test statistic, one can construct a statistical two-sample test for the null hypothesis $H_0: \mathbb{P}_\X = \mathbb{P}_\Y$ against the alternative hypothesis $H_1: \mathbb{P}_\X \neq \mathbb{P}_\Y$ \cite{gretton2012kernel}.

Let $\alpha$ be the significance level of the test, i.e., the maximum allowable probability of falsely rejecting $H_0$ and hence an upper bound on the type-I error.  Given $\alpha$, the threshold $c_\alpha$ for the test statistic can be approximated with a permutation test as follows. 
We first generate $P$ randomly permuted partitions of the set of all realisations $\X \cup \Y$ with sizes commensurate with $(\X, \Y)$, denoted $(\X_p, \Y_p), \, p=1, \ldots, P$.
We then compute $\widehat{\mmd}_u^2(\mathbf{X}_p, \mathbf{Y}_p), \, \forall p$, and sort the results in descending order. Finally,  we select the statistic at position $(1-\alpha) \times P$ as our empirical threshold $\hat{c}_\alpha$.
The null hypothesis $H_0$ is then rejected if $\widehat{\mmd}_u^2(\X, \Y) > \hat{c}_\alpha$. 
For a computationally less expensive (but generally less accurate) option, the inverse cumulative density function of the Gamma distribution can be computed to approximate the null distribution \cite{gretton2009fast}. 

\subsection{\ac{hsic} for non-stationary random processes} \label{hsic}

Let $\P_{\X\Y}$ denote the joint distribution of $\{\X_t\}$ and $\{\Y_t\}$, and let $\mathcal{H}_k$ and $\mathcal{G}_l$ be separable \acp{rkhs} with characteristic kernels $k:\R^{T_\X}\times \R^{T_\X}\rightarrow \R$ and $l:\R^{T_\Y}\times \R^{T_\Y}\rightarrow \R$, respectively. 
\acs{hsic} is then defined as the \ac{mmd} between $\P_{\X\Y}$ and $\P_\X\P_\Y$ ~\cite{gretton2008kernel}:
\begin{align}
\label{eq:HSIC}
    &\hsic(\mathcal{H}_k,  \mathcal{G}_l, \mathbb{P}_{\X \Y}) := \mmd^2(\H_k \otimes \G_l, \P_{\X\Y},\P_\X\P_\Y) \\ 
    & \qquad = \| \mu_{\X\Y} - \mu_{\X} \otimes \mu_{\Y} \|_{\H_k\otimes \G_l}^2 \geq 0, \text{with equality iff} \quad 
    \mathbb{P}_{\X\Y} = \mathbb{P}_{\Y}\P_\Y.
    \nonumber
\end{align}

Here, $\otimes$ denotes the tensor product.
Recall that we assume an equal number of realisations $m$ for both processes, and let  $\Kk, \Lk \in \mathbb{R}^{m \times m}$ be the kernel matrices with entries $k_{ij} = k(\x_i, \x_j)$ and $l_{ij} = l(\y_i, \y_j)$, respectively. 
Given i.i.d. samples $(\X, \Y)$, an unbiased empirical estimator of $\hsic(\mathcal{H}_k, \mathcal{G}_l, \mathbb{P}_{\X \Y})$ is given by~\cite[Theorem 2]{song2012feature}: 
\begin{equation}\label{eq:hsic_u}
    \widehat{\hsic}_u(\mathcal{H}_k, \mathcal{G}_l, \X \Y) = \frac{1}{m(m-3)} \Big[ \text{trace}(\widetilde{\Kk} \widetilde{\Lk}) + \frac{\boldsymbol{1}^\top \widetilde{\Kk} \boldsymbol{1} \, \boldsymbol{1}^\top \widetilde{\Lk} \boldsymbol{1}}{(m-1)(m-2)} - \frac{2}{m-2} \boldsymbol{1}^\top \widetilde{\Kk} \widetilde{\Lk} \boldsymbol{1} \Big] , 
\end{equation}
where $\widetilde{\Kk}=\Kk - \text{diag}(\Kk)$ and $\widetilde{\Lk} = \Lk - \text{diag}(\Lk)$, 
and $\boldsymbol{1}$ is the $m \times 1$ vector of ones. To ease our notation, we henceforth omit the implied $\mathcal{H}_k$ and $\mathcal{G}_l$.

To test $\widehat{\hsic}_u(\X \Y)$ for statistical significance, we define the null hypothesis $H_0: \mathbb{P}_{\X\Y} = \mathbb{P}_\X \mathbb{P}_\Y$ and the alternative $H_1: \mathbb{P}_{\X\Y} \neq \mathbb{P}_\X \mathbb{P}_\Y$. We broadly repeat the procedure outlined in \cref{mmd} by bootstrapping the distribution under $H_0$ via permutations, with the distinction that we only permute the samples $\{\y_i\}_{i=1}^m$, resulting in $\Y_p, p \in [1, P]$, whilst the $\{\x_j\}_{j=1}^m$ are kept unchanged~\cite{gretton2008kernel}. $\widehat{\hsic}_u(\X \Y)$ is then computed for each permutation $(\X, \Y_p)$ and the empirical threshold $\hat{c}_\alpha$ is taken as the statistic at position $(1 - \alpha) \times P$. The null hypothesis $H_0$ is rejected, if $\widehat{\hsic}_u(\X \Y) > \hat{c}_\alpha$.

\subsection{Maximising the test power} \label{maxtestpower}

The power of both \ac{mmd}-based two-sample and \ac{hsic}-based independence tests is prone to decay in high dimensional spaces~\cite{ramdas2015decreasing,reddi2015high}, as in our setting where each measurement point in time is treated as a separate dimension. 
Hence, we describe here how a kernel $k$ can be chosen to maximise the test power, i.e., the probability of correctly rejecting $H_0$ given that it is false.
First, note that under $H_1$ both $\widehat{\mmd}_u^2(\X, \Y)$ \cite[Corollary 16]{gretton2012kernel} and $\widehat{\hsic}_u(\X \Y)$ \cite[Theorem 1]{gretton2008kernel} are asymptotically Gaussian: 
\begin{align}
    \frac{\widehat{\mmd}_u^2(\X, \Y) - \mmd^2(\mathbb{P}_\X, \mathbb{P}_\Y)}{\sqrt{V_m^{\mmd}(\mathbb{P}_\X, \mathbb{P}_\Y)}}
    &\overset{D}{\longrightarrow} \mathcal{N}(0,1)  \\[0.25cm]
    \frac{\widehat{\hsic}_u(\X \Y) - \hsic(\mathbb{P}_{\X \Y})}{\sqrt{V_m^{\hsic}(\mathbb{P}_{\X \Y}})} 
    &\overset{D}{\longrightarrow} \mathcal{N}(0,1),
\end{align}
where $V_m^{\mmd}(\mathbb{P}_\X, \mathbb{P}_\Y)$ and $V_m^{\hsic}(\mathbb{P}_{\X \Y})$ denote the asymptotic variance of $\widehat{\mmd}_u^2(\X, \Y)$ and $\widehat{\hsic}_u(\X \Y)$, respectively \cite[Section 5.5.1 (A)]{Serfling2002Atom}. 

Given a significance level $\alpha$, we define the test thresholds $c^{\mmd}_{\alpha}$ and $c^{\hsic}_{\alpha}$ and reject $H_0$ if $\widehat{\mmd}_u^2(\X, \Y) > c^{\mmd}_\alpha $ or $\widehat{\hsic}_u(\X \Y) > c^{\hsic}_\alpha$. Following \cite{sutherland2016generative}, the test power is defined in terms of $\P_1$, the distributions under $H_1$, with equal sample sizes $m = n$ as:
\begin{align} 
    \label{opt-mmd}
    \P_1 \left( \widehat{\mmd}_u^2(\X, \Y) > \frac{\hat{c}_\alpha^{\mmd}}{m} \right) &\overset{D}{\longrightarrow} 
    \Phi \left( 
    \frac{\mmd^2(\mathbb{P}_\X, \mathbb{P}_\Y)-c_\alpha^{\mmd} / m} {\sqrt{V_m^{\mmd}(\mathbb{P}_\X, \mathbb{P}_\Y)}}  \right)  \\
    \label{opt-hsic}
    \P_1 \left(\widehat{\hsic}_u(\X \Y) > \frac{\hat{c}_\alpha^{\hsic}}{m} \right) &\overset{D}{\longrightarrow} 
    \Phi \left( 
    \frac{\hsic(\mathbb{P}_{\X \Y}) - c_\alpha^{\hsic} / m} {\sqrt{V_m^{\hsic}(\mathbb{P}_{\X \Y})}}  \right) ,
\end{align}
where $\Phi$ is the cumulative density function of the standard Gaussian distribution, and where $\hat{c}_\alpha \rightarrow c_\alpha$ with increasing sample size. To maximise the test power, we maximise the argument of $\Phi$, 
which we approximate by maximising
$\widehat{\mmd}_u^2(\X, \Y) / \sqrt{\hat{V}_m^{\mmd}(\X, \Y)}$ and
minimising $\hat{c}_\alpha^{\mmd} / \Big(m \, \sqrt{\hat{V}_m^{\mmd}(\X, \Y)} \Big)$ for~\eqref{opt-mmd}, and 
%
%
similarly for~\eqref{opt-hsic}. The empirical unbiased variance $\hat{V}_m^{\mmd}(\X, \Y)$ in \eqref{opt-mmd} was derived in \cite{sutherland2016generative}, and we use \cite[Theorem 5]{song2012feature} for $\hat{V}_m^{\text{HSIC}}(\X \Y)$ in \eqref{opt-hsic}.

We perform this optimisation by splitting our samples $(\X, \Y)$ into training and testing sets, of which we take the former to learn the kernel hyper-parameters and the latter to conduct the final hypothesis test with the learnt kernel.

%% file: Section_4.tex
\section{Experimental results on synthetic data} \label{experiments}

To evaluate our proposed tests empirically, we first apply our homogeneity and independence tests to various non-stationary synthetic datasets.
We report test performance using $\hat{\mu}$, the percentage of rejection of the null hypothesis $H_0$, which becomes the test power once $H_0$ is false,
by repeating the experiments on $200$ trials (i.e., $200$ independently generated synthetic datasets).
We provide $95\%$ confidence intervals computed as $\hat{\mu} \pm 1.96 \sqrt{\hat{\mu}(1-\hat{\mu})/200}$.

\subsection{Homogeneity tests with \ac{mmd}} \label{method_mmd}

\paragraph{Setup.}
We evaluate our \ac{mmd}-based homogeneity test against shifts in mean and variance of two non-stationary stochastic processes $\{\X_t\}$ and $\{\Y_t\}$ by establishing if they are correctly accepted or rejected under the null hypothesis $H_0: \P_{\X} = \P_{\Y}$. For ease of comparison, we adopt the experimental protocol of \cite{pomann2016two} and consider two stochastic processes based on a linear mixed effects model.  
We generate independent samples $\X = \{\x_i\}_{i=1}^m$ and $\Y = \{\y_i\}_{i=1}^n$ on
an equally spaced temporal grid of length $T_{\X} = T_{\Y} = T$ in the interval $\mathcal{I} = [0, 1]$,
\begin{align}
    \label{modelX}
    \textstyle
    x_{i,t} &= \mu_\X(t) + \sum_{k=1}^K {\xi_\X}_{i,k} \, \phi_{k}(t) + {\epsilon_\X}_{i,t} 
    \ \ \text{and} \ \
    y_{i,t} = \mu_\Y(t) + \sum_{k=1}^K {\xi_\Y}_{i,k} \, \phi_{k}(t) + {\epsilon_\Y}_{i,t} \, ,
\end{align}
where we set $K=2$ with Fourier basis functions $\phi_{1}(t) = \sqrt{2} \sin(2 \pi t)$ and $\phi_{2}(t) = \sqrt{2} \cos(2 \pi t)$. The coefficients ${\xi_\X}_{i,k}$ and ${\xi_\Y}_{i,k}$ and the additive noises  ${\epsilon_\X}_{i,t}, {\epsilon_\Y}_{i,t}$ are all independent Gaussian-distributed random variables with means and variances specified below. 

We evaluate the test power against varying values of shifts in mean and variance as follows:
\begin{itemize}
    \item \textit{Mean shift:} $\mu_{\X}(t) = t$ and $\mu_{\Y}(t) = t + \delta_{\mu} t^3$. The basis coefficients are sampled as ${\xi_\X}_{i,1}, \, {\xi_\Y}_ {i,1} \sim \mathcal{N}(0, 10)$ and ${\xi_\X}_{i,2}, \, {\xi_\Y}_{i,2} \sim \mathcal{N}(0, 5)$, and the additive noises are sampled as ${\epsilon_\X}_{i,t}, \, {\epsilon_\Y}_{i,t} \sim \mathcal{N}(0, 0.25)$. 
    \item \textit{Variance shift:} We take $\mu_\X(t) = \mu_\Y(t) = 0$, and introduce a shift in variance in the first basis function coefficients via ${\xi_\X}_{i,1} \sim \mathcal{N}(0, 10)$ and ${\xi_\Y}_{i,1} \sim \mathcal{N}(0, 10 + \delta_{\sigma})$. The second coefficients are sampled as ${\xi_\X}_{i,2}, {\xi_\Y}_{i,2} \sim \mathcal{N}(0, 5)$, and the noises as ${\epsilon_\X}_{i,t}, \, {\epsilon_\Y}_{i,t} \sim \mathcal{N}(0, 0.25)$.
\end{itemize}
The coefficients $\delta_\mu$ and $\delta_\sigma$ for mean and variance shifts, respectively, determine the departure from the null hypothesis. Setting $\delta_\mu, \delta_\sigma = 0$ means $H_0$ is true, whereas $\delta_\mu, \delta_\sigma > 0$ means $H_0$ is false. Although this is not a necessity, we set the number of independent samples of $\{\X_t\}$ and $\{\Y_t\}$ to be equal, $m = n$. 
To test for statistical significance, we follow the procedure described in~\cref{mmd} and perform permutation tests of $P=5000$ partitions for varying values of  $\delta_\mu$ and $\delta_\sigma$ and different sample sizes $m=100, 200, 300, 500$.

\begin{figure}[tb]
    \centering
    \begin{minipage}{.47\textwidth}
    \includegraphics[width=\textwidth]{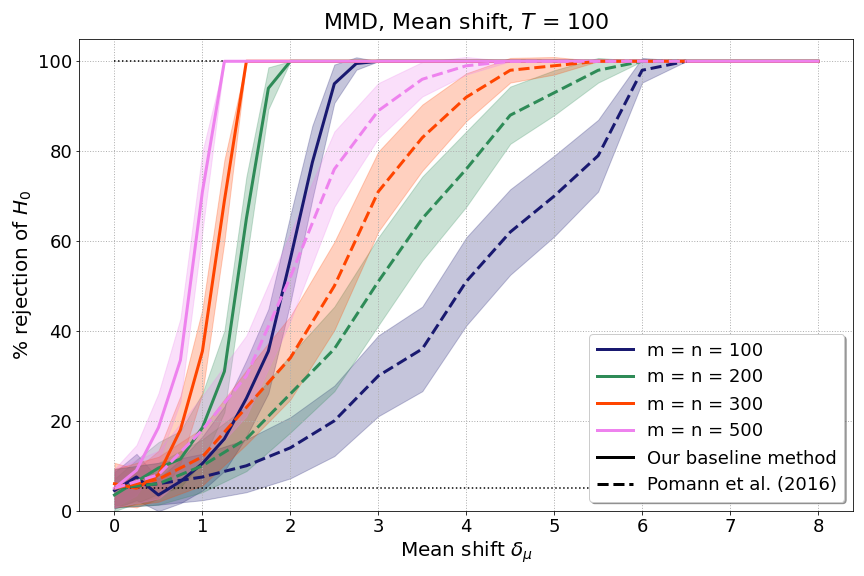}
    \end{minipage}
    \hspace{0.5cm}
    \begin{minipage}{.47\textwidth}
    \includegraphics[width=\textwidth]{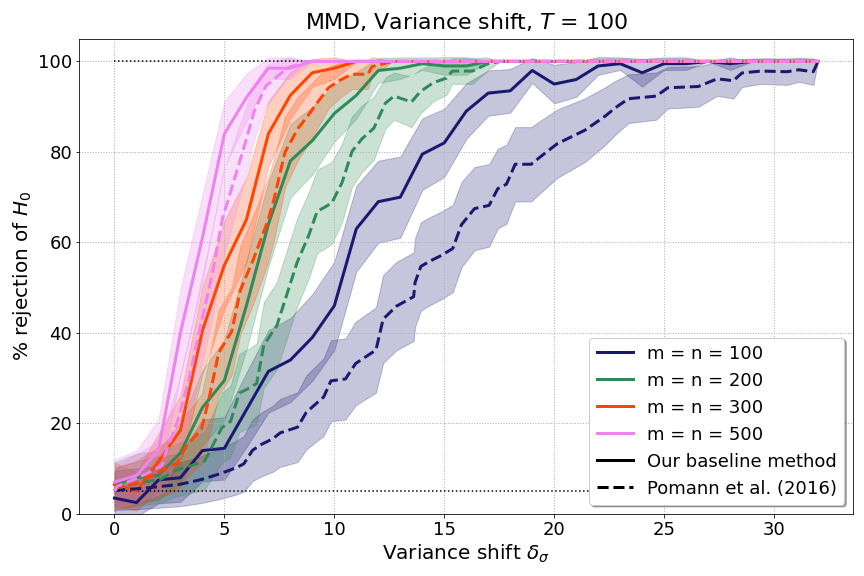}
    \end{minipage}
    \caption{Results of our \ac{mmd}-based homogeneity test for non-stationary random processes: Percentage of rejected $H_0$ as mean shift \textit{(left)} and variance shift \textit{(right)} are varied. 
    Our baseline method 
    (solid lines) is compared to \cite{pomann2016two} (dashed lines) for different sample sizes $m = n=100, 200, 300, 500$ and $T = 100$ discrete time points.}
    \label{fig:mean_var_shift_100}
\end{figure}

\paragraph{Baseline results without test power optimisation.}

Our baseline results are obtained with a Gaussian kernel $k(x, y) = \exp{(-\|x-y\|^2/\sigma^2)}$ with bandwidth $\sigma$ equal to the median distance between observations of the aggregated samples. Figure \ref{fig:mean_var_shift_100} shows how our method (solid lines) compares to \cite{pomann2016two} (dashed lines) for $T = 100$ discrete time points. For all sample sizes, the type-I error rate lies at or below the allowable probability of false rejection $\alpha$, and our method significantly outperforms \cite{pomann2016two} for nearly all levels of mean and variance shifts. Both shifts become easier to detect for larger sample sizes. Particularly strong improvements are achieved for mean shifts: our method makes no type-II errors for $\delta_\mu \geq 3$ on $m=100$ samples, whereas \cite{pomann2016two} only reach such performance with $m=500$ samples and $\delta_\mu \geq 4.5$. We obtain similar test power results (see Appendix~\ref{results_coarse}) for coarser realisations with $T = 5, 10, 25, 50$ over the same interval $\mathcal{I} = [0,1]$.

\paragraph{Results of the optimised test.} 

\begin{figure}[tb]
    \centering
    \begin{minipage}{.47\textwidth}
    \includegraphics[width=\textwidth]{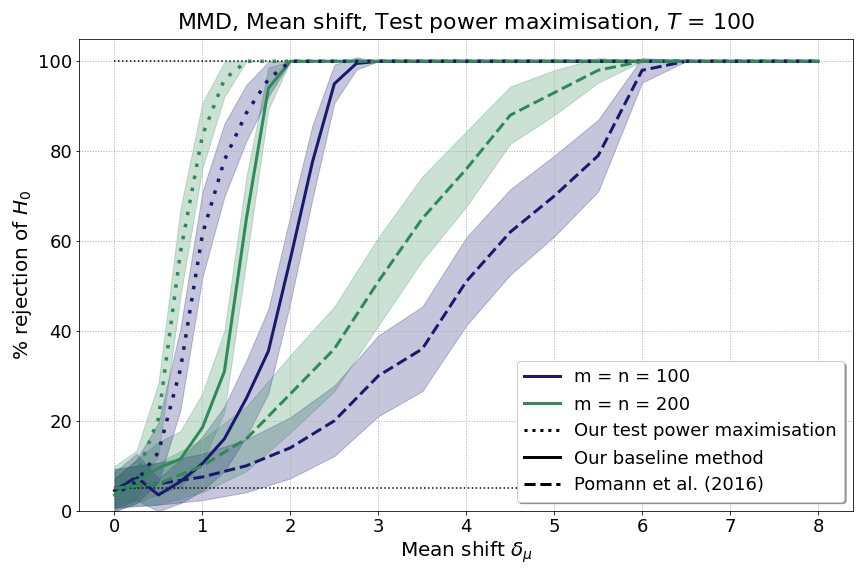}
    \end{minipage}
    \hspace{0.5cm}
    \begin{minipage}{.47\textwidth}
    \includegraphics[width=\textwidth]{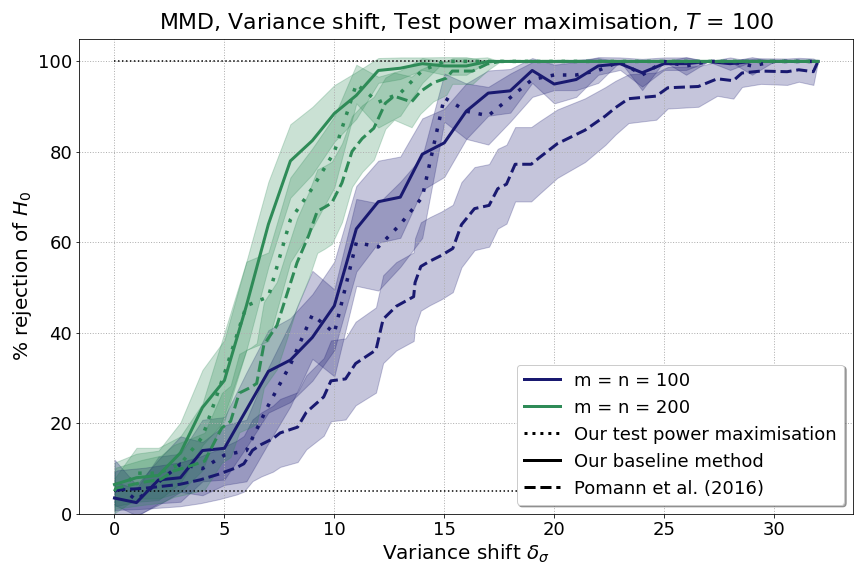}
    \end{minipage}
    \caption{Results of homogeneity test with optimising for test power: Percentage of rejected $H_0$ for mean shift \textit{(left)} and variance shift \textit{(right)} for sample sizes $m = n = 100, 200$ and $T = 100$ discrete time points. Our optimised test power method (dotted lines) is compared to our baseline method (solid lines) and \cite{pomann2016two} (dashed lines).}
    \label{fig:mean_var_shift_100_max}
\end{figure}

Next, we apply the method described in~\cref{maxtestpower} to maximise the test power. 
Specifically, we search for the Gaussian kernel bandwidth $\sigma$ (over spaces defined in Table~\ref{search_spaces_MMD} in Appendix~\ref{delta_search}), that maximises the argument of $\Phi$ in our approximations of \eqref{opt-mmd} on our training samples. For demonstrative purposes, we choose to split our dataset equally into training and testing sets although other ratios may lead to higher test power.
Figure~\ref{fig:mean_var_shift_100_max} shows the results of the optimised test (dotted lines) against the baseline results (solid lines) and the results of \cite{pomann2016two} (dashed lines) for $m=100$ and $m=200$ samples and $T = 100$ discrete points in time. We find that the test power is significantly improved by our optimisation for the detection of mean shifts. For instance, test power rises fourfold for $\delta_{\mu} = 1$ and $m = 200$ compared to our baseline method. Furthermore, we have no type-II errors once $\delta_{\mu} \geq 2$ for $m = 100$, as compared to $\delta_\mu \geq 3$ for our baseline test and $\delta_\mu \geq 6.5$ for \cite{pomann2016two}. 
In its current form, however, our optimisation does not yield higher test power for the detection of variance shifts, a fact that we discuss in~\cref{conclusion}.

\subsection{Independence tests with \ac{hsic}}  \label{experiments_hsic}

\paragraph{Setup.}
To test for independence, the null hypothesis is $H_0: \mathbb{P}_{\mathbf{XY}} = \mathbb{P}_{\X} \mathbb{P}_{\Y}$. We assume we observe measurements $x_{i,t}$ and $y_{i,t}$ over temporal grids of length $T_\X$ and $T_\Y$ in the interval $\mathcal{I} = [0, 1]$, respectively. To measure type-I and type-II error rates, we use the following experimental protocols, partly adopted from~\cite{zhang2018large} and \cite{gretton2008kernel,gretton2005kernel}:  
\begin{itemize}
    \item \textit{Linear dependence:} $\X$ is generated as in~\eqref{modelX} with $\mu_\X(t) = t$, basis coefficients ${\xi_\X}_{i,1} \sim \mathcal{N}(0, 10)$, ${\xi_\X}_{i,2} \sim \mathcal{N}(0, 5)$, and noise ${\epsilon_\X}_{i,t} \sim \mathcal{N}(0, 0.25)$. The samples of the second process are $\Y = \{ x_{i,1} + \epsilon_i\}_{i=1}^m$ where
    $\epsilon_i \sim \mathcal{N}(0,1)$, as in~\cite{zhang2018large}.
    %
    %
    \item \textit{Dependence through a shared coefficient:} $\X$ and $\Y$ are generated as in~\eqref{modelX} with $\mu_\X(t) = \mu_\Y(t) = t$ and independently sampled ${\xi_\X}_{i,1}$, ${\xi_\Y}_{i,1}$,  ${\epsilon_\X}_{i,t}$, ${\epsilon_\Y}_{i,t}$ as in the mean shift experiments of~\cref{method_mmd}, but where the stochastic processes now share the second basis function coefficient: ${\xi_\X}_{i,2} = {\xi_\Y}_{i,2}$. 
    %
    \item \textit{Dependence through rotation:} We start by generating independent $\X^{(0)}$ and $\Y^{(0)}$ as in~\eqref{modelX} with $\mu_\X(t) = \mu_\Y(t) = t$ and ${\epsilon_\X}_{i,t}, {\epsilon_\Y}_{i,t} \sim \mathcal{N}(0, 0.25)$, but with ${\xi_\X}_{i,k}$ and ${\xi_\Y}_{i,k}$ drawn from: (i) student-t, (ii) uniform, or (iii) exponential distributions~\cite[Table 3]{gretton2005kernel}. 
    We next multiply $(\X^{(0)}, \Y^{(0)})$
    by a $2\times2$ rotation matrix $R(\theta)$ with $\theta \in [0,\pi/4]$ to generate new rotated samples $(\X,\Y)$, which we then test for independence. Clearly, for $\theta = 0$ our samples $(\X,\Y)$ are independent and as $\theta$ is increased their dependence becomes easier to detect (see \cite[Section 4]{gretton2008kernel} and Figure~\ref{fig:rotations_illustrations} for implementation details).
\end{itemize} 
Statistical significance is computed using $P=5000$ permutations of $\Y$ whilst $\X$ is kept fixed to approximate the distribution under $H_0$. Test power is calculated for varying $T = [5, 10, 25, 50, 100]$ and different sample sizes $m = n$.  

\paragraph{Baseline results without test power optimisation.} 

\begin{figure}[tb]
    \centering
    \begin{minipage}{.47\textwidth}
    \includegraphics[width=\textwidth]{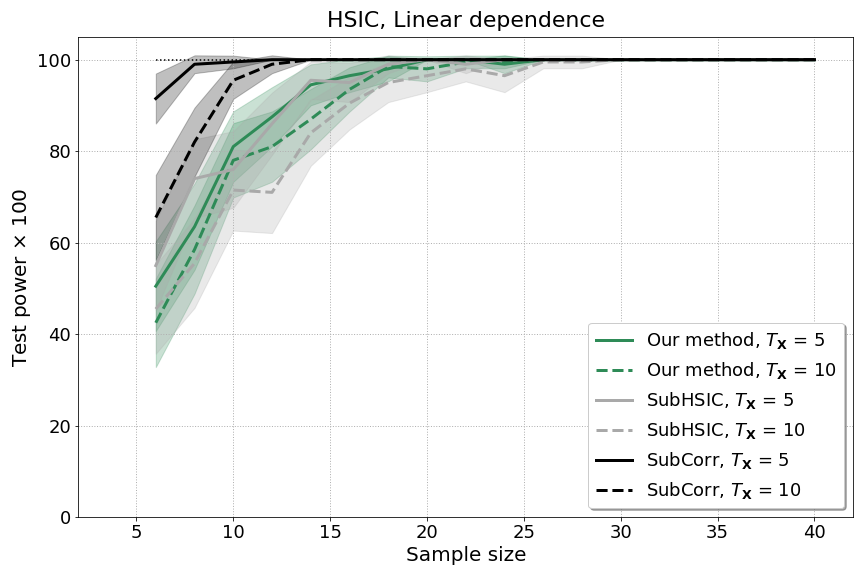}
    \end{minipage}
    \hspace{0.5cm}
    \begin{minipage}{.47\textwidth}
    \includegraphics[width=\textwidth]{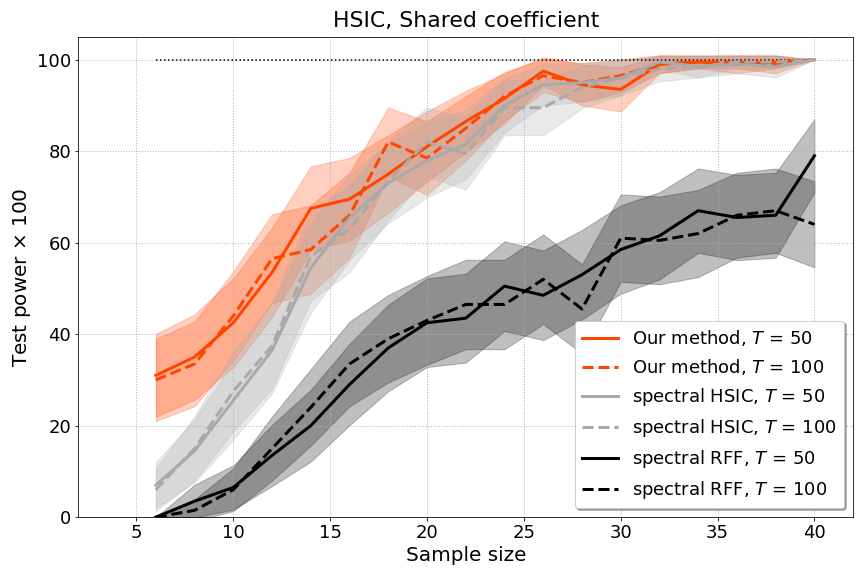}
    \end{minipage}
    \caption{ 
    Results of the \ac{hsic}-based independence test: Test power for linear dependence \textit{(left)} and  dependence through shared coefficients 
    \textit{(right)} as sample size is varied for various numbers of time points. For the linear dependence, we compare our baseline results to SubCorr and SubHSIC; for the shared coefficient, we compare against two spectral approximations \cite[Section 5.1]{zhang2018large}.}
    \label{fig:dep_linear_coeff}
\end{figure}

Our baseline results are computed using a Gaussian kernel with $\sigma$ equal to the median distance between measurements in the corresponding sample.
Figure~\ref{fig:dep_linear_coeff} \textit{(left)} shows the results of our test on the linear dependence experiments, which demonstrate, due to $T_\Y = 1$, how dependencies between individual points in time and an entire time series can be detected.
We compare our method to: 
(i) a statistic explicitly aimed at linear dependence, $\text{SubCorr} = \frac{1}{T_\X} \sum_{t=1}^{T_\X} \text{Corr}(\{x_{i,t}\}_{i=1}^m, \Y)$, where $\text{Corr}(\cdot, \cdot)$ is the Pearson correlation coefficient; 
and (ii) $\text{SubHSIC} = \frac{1}{T_\X} \sum_{t=1}^{T_\X} \widehat{\hsic}_u (\{x_{i,t}\}_{i=1}^m, \Y)$. For both of these methods, the distribution under $H_0$ is also approximated via permutations. 
We find that SubCorr outperforms the other methods in experiments with sample sizes $m < 20$, and SubHSIC achieves comparable results to our method. 
The results for $T_\X = [25, 50, 100]$ (see Appendix~\ref{results_coarse}) are similar.

Figure~\ref{fig:dep_linear_coeff} \textit{(right)} displays the power of our independence test for the case of 
dependent samples through a shared coefficient 
for varying sample sizes $m$ and measurements $T$. 
We compare our results to two spectral methods \cite{zhang2018large} that approximate the distribution under $H_0$ using eigenvalues of the centred kernel matrices of $\X$ and $\Y$: spectral \ac{hsic} uses the unbiased estimator \eqref{eq:hsic_u} as the test statistic with the eigenvalue-based null distribution; and spectral \ac{rff} uses a test statistic induced by a number of \acp{rff} (set here to $10$) that approximate the kernel matrices of $\X$ and $\Y$. Our method and spectral \ac{hsic} achieve $20-50\%$ improvement in test power compared to spectral \ac{rff}.
For small numbers of samples ($m<15$), our method outperforms spectral \ac{hsic}, which converges to the performance of our method with increasing sample size, as we would expect it \cite[Theorem 1]{gretton2009fast}. 

\begin{figure}[b]
    \centering
    \begin{minipage}{.32\textwidth}
    \includegraphics[width=\textwidth]{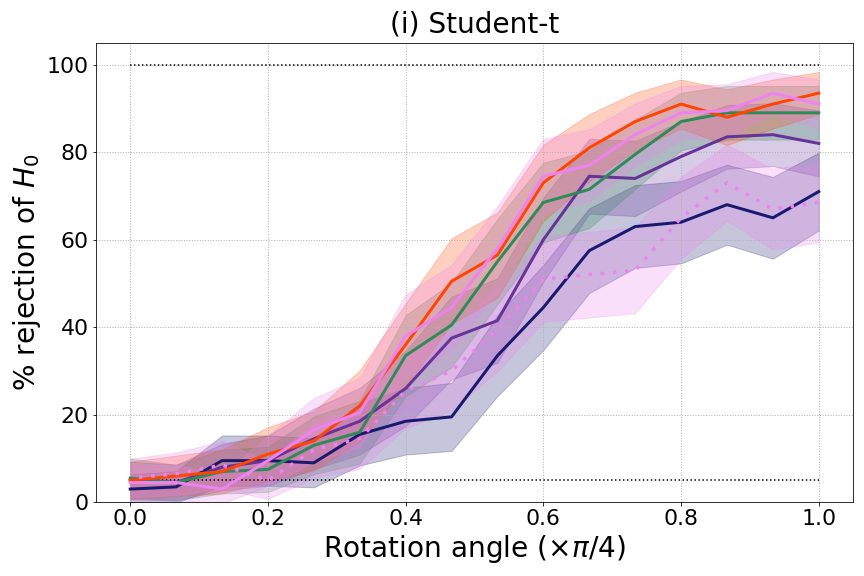}
    \end{minipage}
    \begin{minipage}{.32\textwidth}
    \includegraphics[width=\textwidth]{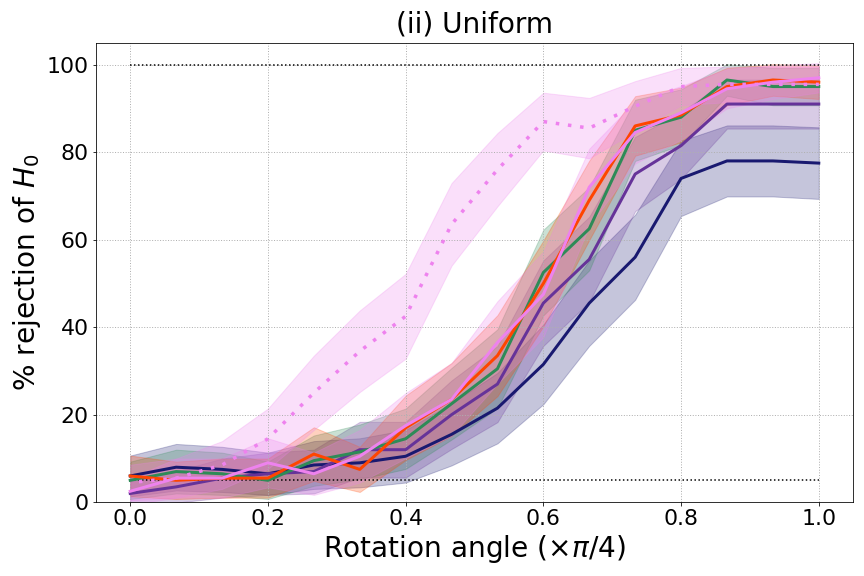}
    \end{minipage}
    \begin{minipage}{.32\textwidth}
    \includegraphics[width=\textwidth]{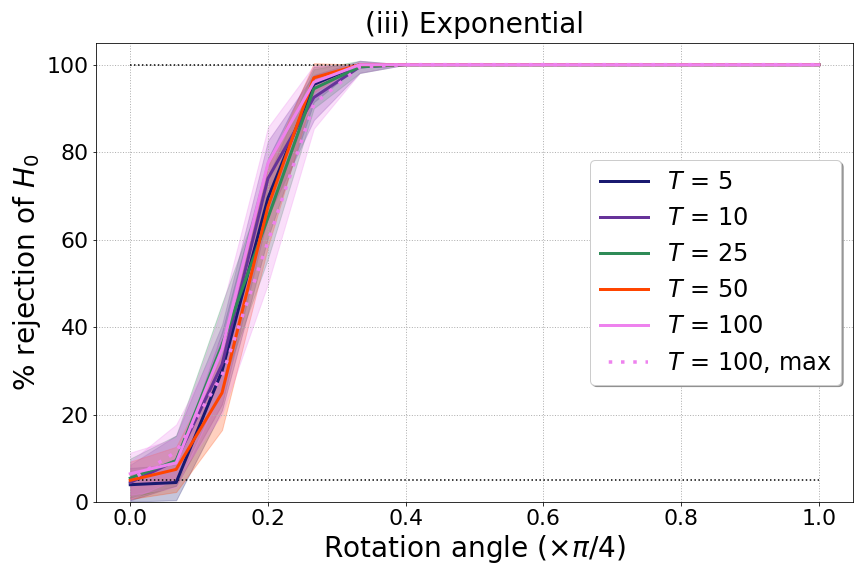}
    \end{minipage}
    \caption{
    Results of the \ac{hsic}-based independence test: 
    Percentage of rejected $H_0$ in rotation dependence experiments for different number of discrete time points $T$ and coefficients ${\xi_\X}_{i,k}$ and ${\xi_\Y}_{i,k}$ drawn from three distributions: (i) student-t, (ii) uniform, and (iii) exponential (see \ref{tab:dist_coeff}). The sample size is $m = 200$. The violet dotted lines are the results of our test power maximisation.}
    \label{fig:dep_rotation}
\end{figure}

Figure~\ref{fig:dep_rotation} shows the rotation dependence experiments, where $\theta = 0$ corresponds to the null hypothesis (independence) and $\theta > 0$ to the alternative. The distribution hyper-parameters for ${\xi_\X}_{i,k}$ and ${\xi_\Y}_{i,k}$ are detailed in Appendix~\ref{tab:dist_coeff}, and we set $T_\X = T_\Y = T$, although equality is not required. As expected, dependence is easier to detect with increasing $\theta$. We observe that denser temporal measurements do not result in enhanced test power. Note that the test power is highly dependent on the distribution of the coefficients of the basis functions ${\xi_\X}_{i,k}$,  ${\xi_\Y}_{i,k}$. 

\paragraph{Results of the optimised test.} 

The test power maximisation was applied to the rotation dependence experiments 
by searching for optimal Gaussian kernel bandwidths $\sigma_\X$ and $\sigma_\Y$ over pre-defined intervals (specified in Appendix~\ref{delta_search}). 
Figure~\ref{fig:dep_rotation} shows that the test power is improved when the basis function coefficients 
are drawn from uniform distributions. In this case, the percentage of rejected $H_0$ is $20-40\%$ higher 
for $\theta$ between $0.2$ and $0.75 \times \pi / 4$, but it levels off at $95\%$ once $\theta \geq 0.75 \times \pi/4$, which is the same level achieved by our baseline method for $\theta \geq 0.85 \times \pi/4$. With our current test-train split, our optimised test does not improve the test power if the basis function coefficients ${\xi_\X}_{i,k}$ and ${\xi_\Y}_{i,k}$ are drawn from student-t or exponential distributions.

\section{Application to a socio-economic dataset} \label{sdgs}

As a further illustration, we apply our method to the United Nations' socio-economic~\acp{sdg} (see Appendix~\ref{sdg data} for details). Specifically, we investigate whether some so-called Targets of the 17 \acp{sdg} have been homogeneous over the last $20$ years across low- and high-income countries, and whether certain \acp{sdg} in African countries exhibit dependence over the same period. In both setting, we assume countries are independent.

For our homogeneity tests, we classify countries into low- and high-income according to \cite{worldbank2020}. We use temporal data of $76$ Targets for which \cite{WBdata} provides data collected over the last $T = 20$ years for $m = 30$ low-income countries and $n = 55$ high-income countries.
Applying our baseline method without test power optimisation, we find that, out of the $76$ Targets we have data available for, only $38$ have had homogeneous trajectories in low- and high-income countries.
For instance, whereas the `death rate due to road traffic injuries' (Target~3.6) has been homogeneous between these two groups, the `fight the epidemics of AIDS, tuberculosis, malaria and others' (Target~3.3) has not been homogeneous in low- and high-income countries. 

For our independence tests, we consider temporal data from $m = n = 49$ African countries over $T=20$ years, and test any two Targets for pairwise independence. Of the total $2850$ possible pairwise combinations, the null hypothesis of independence is rejected for $357$. As an illustration, we examine the dependencies of `implementation of national social protection systems' (Target~1.3) with `economic growth' (Target~8.1) and the `proportion of informally employed workers' (Target~8.3).
Applying our baseline method, we accept the null hypothesis of independence between Target~1.3 and 8.1, i.e., we find that the `implementation of national social protection systems' has been independent of economic growth. In contrast, we find that Target~1.3 has been dependent on the `proportion of informally employed workers' (Target~8.3).

%% file: Section_5.tex
\section{Discussion and conclusion} \label{conclusion}

Building on ideas from functional data analysis, we have presented approaches to testing for homogeneity and independence between two non-stationary random processes with the kernel-based statistics \ac{mmd} and \ac{hsic}. 
We view independent realisations of the underlying processes as samples from multivariate probability distributions to which \ac{mmd} and \ac{hsic} can be applied. Our tests  are shown to outperform current state-of-the-art methods in a range of experiments. 
Furthermore, we optimise the test power over the choice of kernel and achieve improved results in most settings.
However, we also observe that our optimisation procedure does not always yield an increase in test power. We leave the investigation of this behaviour open for future research with the possibility of defining search spaces and step sizes over kernel hyper-parameters differently, or of choosing a gradient-based approach for optimisation \cite{sutherland2016generative}.
Our results show that small sample sizes of less than $40$ independent realisations can already achieve high test power, and that denser measurements over the same time period do not necessarily lead to enhanced test power. 

The proposed tests can be of interest in many areas where non-stationary and non-linear multivariate temporal datasets constitute the norm, as illustrated by our application to test for homogeneity and independence between the United Nations'~\aclp{sdg} measured in different countries over the last $20$ years.